\documentclass[fleqn,twoside]{article}

\usepackage{espcrc2}

\usepackage{graphicx}

\usepackage[figuresright]{rotating}

\newcommand{\AmS}{{\protect\the\textfont2

  A\kern-.1667em\lower.5ex\hbox{M}\kern-.125emS}}

\title{Exclusive Measurements of $pp\rightarrow d\;\pi^+\pi^0$ :
  Double-Pionic Fusion without $ABC$ Effect\\
(CELSIUS/WASA Collaboration)}

\author{F.~Kren\address[PIT]{Physikalisches Institut der Universit\"at
  T\"ubingen, D-72076 T\"ubingen, Germany},
M.~Bashkanov\addressmark[PIT],
 D.~Bogoslawsky\address[JINR]{Joint Institute for Nuclear Research, Dubna,
  Russia},
H.~Cal\'en\address[SL]{The Svedberg Laboratory, Uppsala, Sweden},
H.~Clement\addressmark[PIT],
L.~Demiroers\address[HU]{Hamburg University, Hamburg, Germany},
C.~Ekstr\"om\addressmark[SL],
K.~Fransson\addressmark[SL],
J.~Greiff\addressmark[SL],
L.~Gustafsson\address[UU]{Uppsala University, Uppsala,Sweden},
B.~H\"oistad\addressmark[UU],
G.~Ivanov\addressmark[JINR],
M.~Jacewicz\addressmark[UU],
E.~Jiganov\addressmark[JINR],
T.~Johansson\addressmark[UU],
O.~Khakimova\addressmark[PIT],
S.~Keleta\addressmark[UU],
I.~Koch\addressmark[UU],
S.~Kullander\addressmark[UU],
A.~Kup\'s\'c\addressmark[SL],
P.~Marciniewski\addressmark[SL],
R.~Meier\addressmark[PIT],
B.~Morosov\addressmark[JINR],
C.~Pauly\address[FJ]{Forschungszentrum J\"ulich, Germany},
H.~Petr\'en\addressmark[UU], 
Y.~Petukhov\addressmark[JINR],
A.~Povtorejko\addressmark[JINR],
R.J.M.Y.~Ruber\addressmark[SL],
K.~Sch\"onning\addressmark[UU],
W.~Scobel\addressmark[HU],
T.~Skorodko\addressmark[PIT],
B.~Shwartz\address[BINP]{Budker Institute of Nuclear Physics, Novosibirsk,
  Russia},
J.~Stepaniak\address[SINS]{Soltan Institute of Nuclear Studies, Warsaw and
  Lodz, Poland},
P.~Th\"orngren-Engblom\addressmark[UU],
V.~Tikhomirov\addressmark[JINR],
G.J.~Wagner\addressmark[PIT], 
M.~Wolke\addressmark[UU],
A.~Yamamoto\address[HEARO]{High Energy Accelerator Research Organization,
  Tsukuba, Japan},
 J.~Zabierowski\addressmark[SINS],
and
J.~Zlomanczuk\addressmark[UU]}

\begin{document}

\begin{abstract}

Exclusive measurements of the reaction $pp\rightarrow d \pi^+ \pi^0$
 have been carried out at $T_p = 1.1$ GeV at the CELSIUS storage ring using
 the WASA detector. The isovector $\pi^+\pi^0$ channel exhibits no enhancement
 at low invariant $\pi\pi$ masses, i. e. no ABC effect. Therefore this most
 basic isovector double-pionic fusion reaction  qualifies as an ideal test
 case for the conventional $t$-channel $\Delta\Delta$ excitation
 process. Indeed, the obtained differential distributions reveal the
 conventional $t$-channel $\Delta\Delta$ mechanism as the appropriate reaction
 process, which also accounts for the observed energy dependence of the total
 cross section. This is an update of a previously published version -- see
 important note at the end of the article.

\vspace{1pc}

\end{abstract}


\maketitle

\section{Introduction}

Double-pionic fusion has been an intriguing reaction all the time since
Abashian, Booth and Crowe\cite{abc} discovered the so-called ABC effect nearly
50 years ago.
The ABC effect stands for an unexpected low-mass enhancement in the spectrum
of the invariant $\pi\pi$ masses $M_{\pi\pi}$. It is named after the initials
of the authors of the first publications on this effect observed in the double
pionic fusion of deuterons and protons to $^3$He. Follow-up
experiments \cite{ban,wur,col} revealed this effect to show up in cases, when
the two-pion production process leads to a bound nuclear system. Since the
effect was not observed in $pd \to ^3$H$\pi^+\pi^0$ data \cite{abc,ban}, this
finding has been taken as evidence that the ABC effect might be restricted to
the scalar-isoscalar $\pi\pi$ channel ($\sigma$ channel). With
the exception of low-statistics bubble-chamber measurements all early
experiments conducted on this issue have been inclusive measurements carried
out preferentially with single-arm magnetic spectrographs for the detection
of the fused nuclei.

\begin{figure} [t]
\begin{center}
\includegraphics[width=0.2\textwidth]{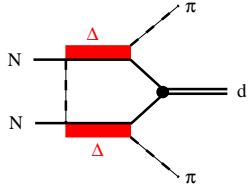}
\caption{ 
  Graph of the double-pionic fusion process via $t$-channel $\Delta\Delta$
  excitation in the intermediate state.
}
\label{fig1}
\end{center}
\end{figure}

Theoretically the ABC effect has been interpreted
\cite{ban,wur,col,ris,anj,gar,alv}  by $t$-channel $\Delta\Delta$ excitation 
in the course of the reaction process (see Fig. 1) leading to both a low-mass
and a 
high-mass enhancement in isoscalar $M_{\pi\pi}$ spectra. In fact, the
missing momentum spectra from inclusive measurements have been in support of
such predictions. 

However, recent exclusive measurements of the isoscalar double-pionic fusion
reactions $pn \to d\pi^0\pi^0$, $pd \to ^3$He$\pi\pi$ and $dd \to ^4$He$\pi\pi$
\cite{bash,MBa,hcl,sk,panic} covering practically the full reaction phase space
exhibit no significant high-mass enhancement, only a very pronounced
low-mass enhancement. Also the new data on the most basic
double-pionic fusion reaction for the ABC effect -- the reaction $pn \to
d\pi^0\pi^0$ -- point to an isoscalar $s$-channel resonance as origin of the
ABC-effect, which couples to $pn$ and $\Delta\Delta$ channels, has a mass of
about 90 MeV below the nominal $\Delta\Delta$ threshold of $\sqrt{s} = 2
m_\Delta$ and a width of only $\approx$ 50 MeV \cite{panic}. The latter is
much smaller than that expected from a conventional $\Delta\Delta$ system,
which has a width of about twice the $\Delta$ width. i.e. $\approx$ 230 MeV.  

If this interpretation is true, then double-pionic fusion leading to isovector
final states is not 
expected to exhibit the ABC-effect. Rather one expects agreement with the
conventional $\Delta\Delta$ t-channel excitation (Fig. 1). Also since due to
Bose symmetry the isovector $\pi^+\pi^0$ system must be in odd relative angular
momentum, i. e. in practice in relative p-wave to each other, there should be
no low-mass enhancement in the  $M_{\pi^+\pi^0}$ spectrum. Hence the $pp
\rightarrow d\pi^+\pi^0$ reaction represents an important test case both for
the current understanding of the ABC effect as a purely isoscalar phenomenon and
for the interpretation of all other double-pionic fusion processes as a
conventional $t$-channel $\Delta\Delta$ process. We note that this
mechanism by itself is of key interest,  since it constitutes the basic
double excitation process in the NN system. Surprisingly it has not yet been 
tested in detail by exclusive and kinematically complete measurements. In a
recent  study \cite{iso} of total cross sections of the non-fusion
channels $pp \to NN\pi\pi$ it has been shown that Roper and
$\Delta\Delta$ excitations are the predominant two-pion production processes.
For a detailed study of the latter the isovector $\pi\pi$ channel is
particularly favorable, since there  the competing Roper excitation is
heavily suppressed \cite {iso,luis}. 

Experiments devoted to the isovector double-pionic fusion are very sparse. For
the $pd \to ^3$H$\pi^+\pi^0$ reaction there are just two single-arm magnetic
spectrometer measurements providing inclusive momentum spectra at just one or
two angle settings \cite{abc,ban}. Experimental information on the 
$pp \rightarrow d\pi^+\pi^0$ reaction is available solely from bubble chamber
measurements \cite{shim,bys} of very low statistics with no information on
differential observables. Hence an exclusive measurement of solid statistics
and covering most of the relevant phase space appears to be crucial
for the most basic isovector double-pionic fusion reaction, 
in order to settle the issue of the ABC effect in the isovector $\pi\pi$
channel as well as to have a clear-cut test case for the conventional
$t$-channel $\Delta\Delta$ process.

\section{Experiment}

In order to investigate this particular issue with detailed experimental
information we have carried out exclusive
measurements of the $pp \rightarrow d\pi^+\pi^0$ reaction at $T_p$ = 1.1 GeV,
i. e. in the energy region of ABC effect and $\Delta\Delta$ excitation. The
experiment was performed at the CELSIUS ring at 
Uppsala using the 4$\pi$ WASA detector setup including the pellet target
system \cite{barg}.

\begin{figure} [t]
\begin{center}
\includegraphics[width=0.23\textwidth]{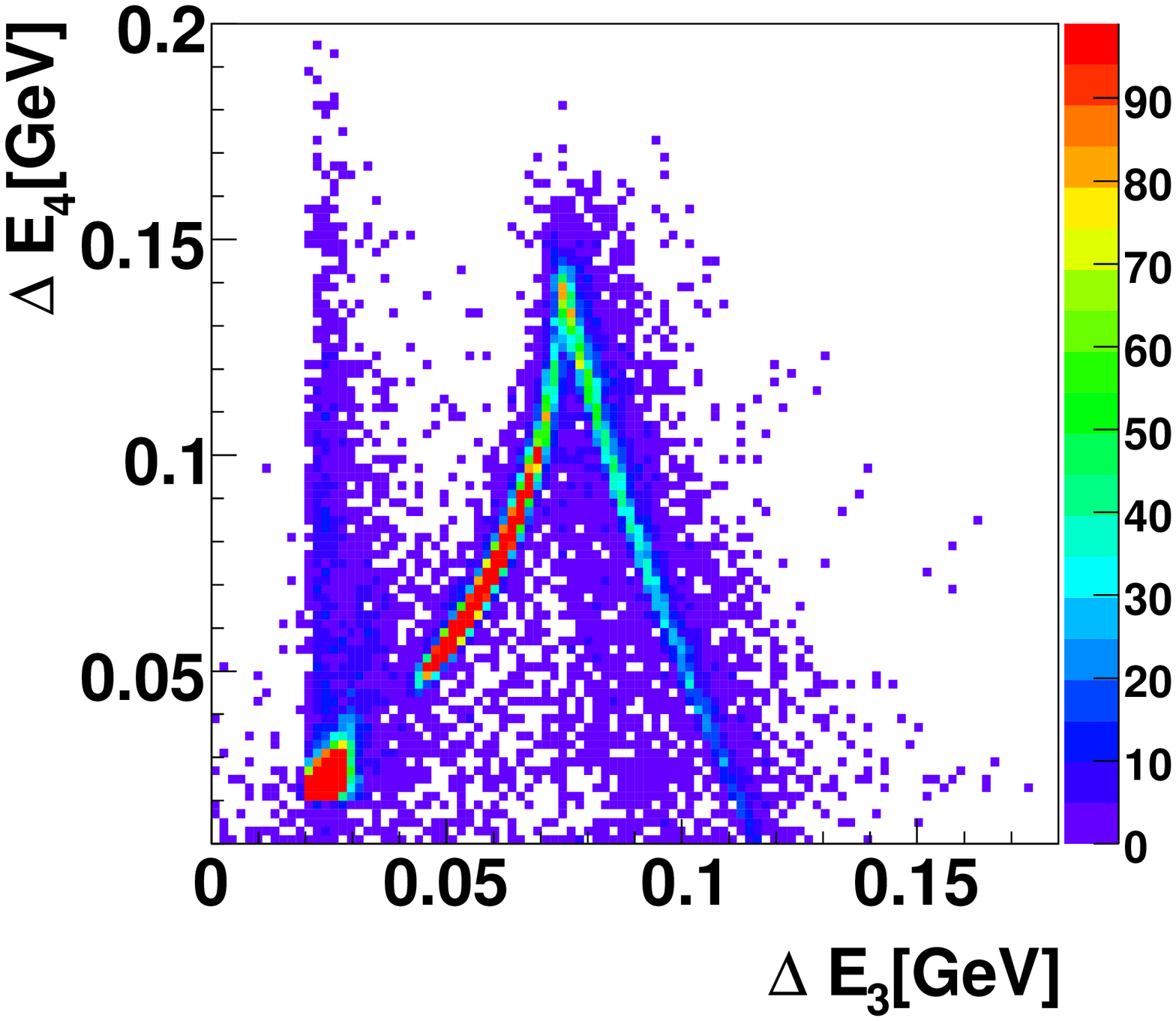}
\includegraphics[width=0.23\textwidth]{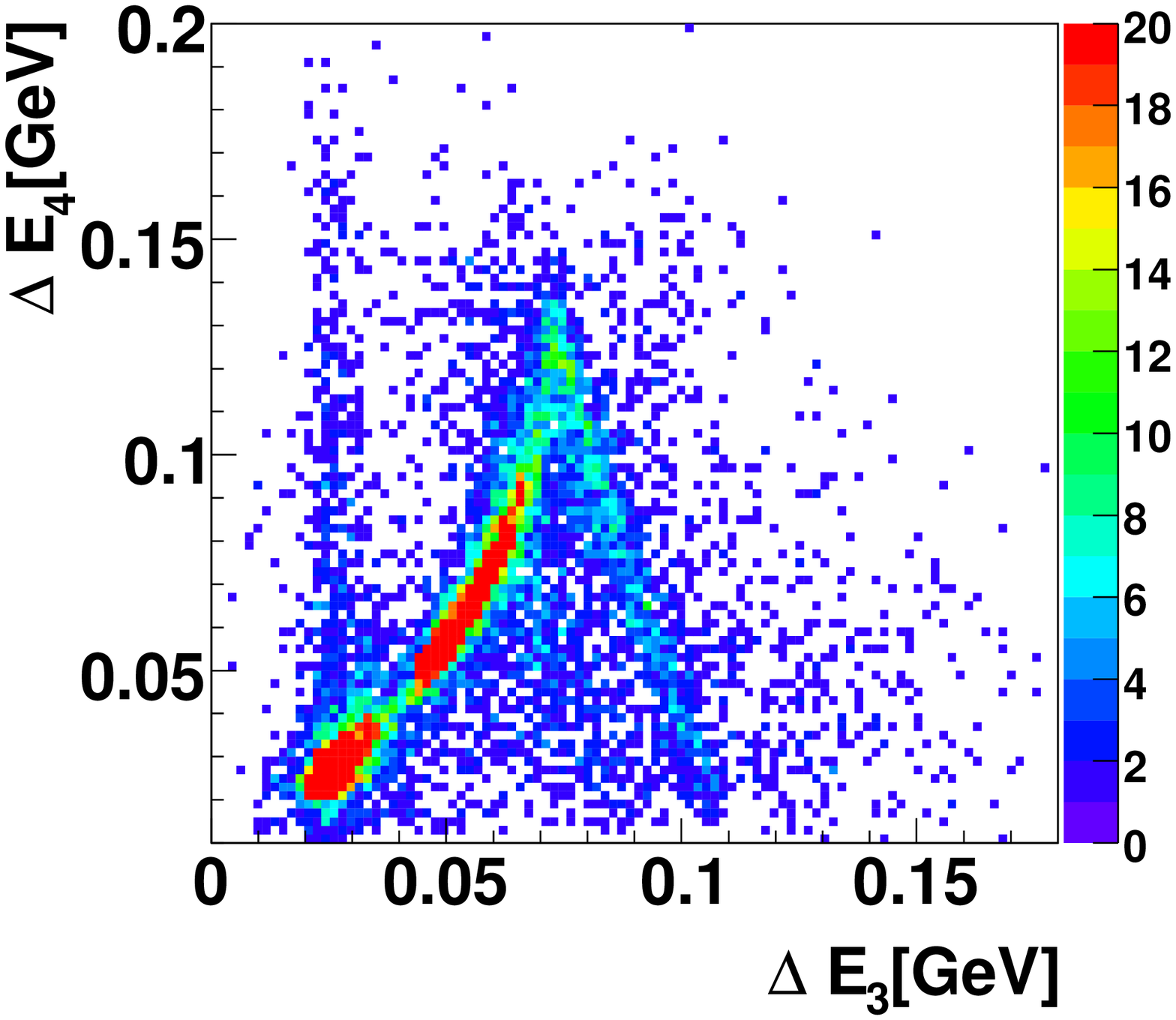}
\caption{ 
  Energy deposit in the third plane of the forward range hodoscope versus
  that in the fourth plane after application of neural net constraints due to
  the other $\Delta$E-E combinations in the multi-layer forward detector. The
  triangle-like pattern shows deuterons, which stop (right wing) or punch
  through (left wing) the $4^{th}$ layer of the forward range hodoscope. Left:
  Monte-Carlo simulation, right: data. 
}
\label{fig1}
\end{center}
\end{figure}


Deuterons and $\pi^+$ particles were detected in the forward detector and
identified by the  
$\Delta$E-E technique using corresponding informations from the quirl and
range hodoscopes, respectively. Since both quirl and range hodoscope consist
of several detector layers each, the energy loss method can be applied for the
particle identification. For this $\Delta$E-E method all possible two-layer
combinations have been used. Most efficiently this can be done by a neural net,
which has been trained by Monte Carlo simulations of the detector performance
\cite{bash,MB}. Fig. 2 shows as an example the combination energy loss
$\Delta E3$ in the $3^{rd}$ hodoscope layer versus the energy loss $\Delta E4$
in the $4^{th}$ layer after application of constraints by neural
net techniques, which clean the spectrum from the large proton
background. Aside from minimum ionizing particles showing up at the lower left
corner of Fig. 2 the deuteron band shows up clearly. The gammas from $\pi^0$
decay were detected in the central detector. 

Since in the experiment $d$, $\pi^+$ and $\pi^0$ have been identified and
since their kinetic energies and scattering angles were measured, the full
four-momentum information about all ejectiles of an $pp \to d\pi^+\pi^0$ event
has been obtained. Together with the additional constraint that the two
detected gammas have to give the $\pi^0$ mass we have thus 5 overconstraints
for the subsequent kinematical fit.  

Acceptance and efficiency corrections have been performed with Monte Carlo
simulations of the detector performance in an iterative procedure starting
with the use of pure phase space distributions and ending with a model, which
provides a satisfactory description of the data (solid lines in Figs. 4 - 6)
assuring thus internal consistency of the procedure.

The absolute cross section was determined by normalization relative to
the $pp \to pp\pi^0$ reaction, which was measured simultaneously with the
same hardware trigger. As a result we obtain a total cross section of 0.092(15)
mb. The uncertainty mainly originates from the large scatter in the literature
values \cite{bys} for the total cross section of the $pp \to pp\pi^0$ reaction 
in the energy region of interest.

\begin{figure} [t]
\begin{center}
\includegraphics[width=0.23\textwidth]{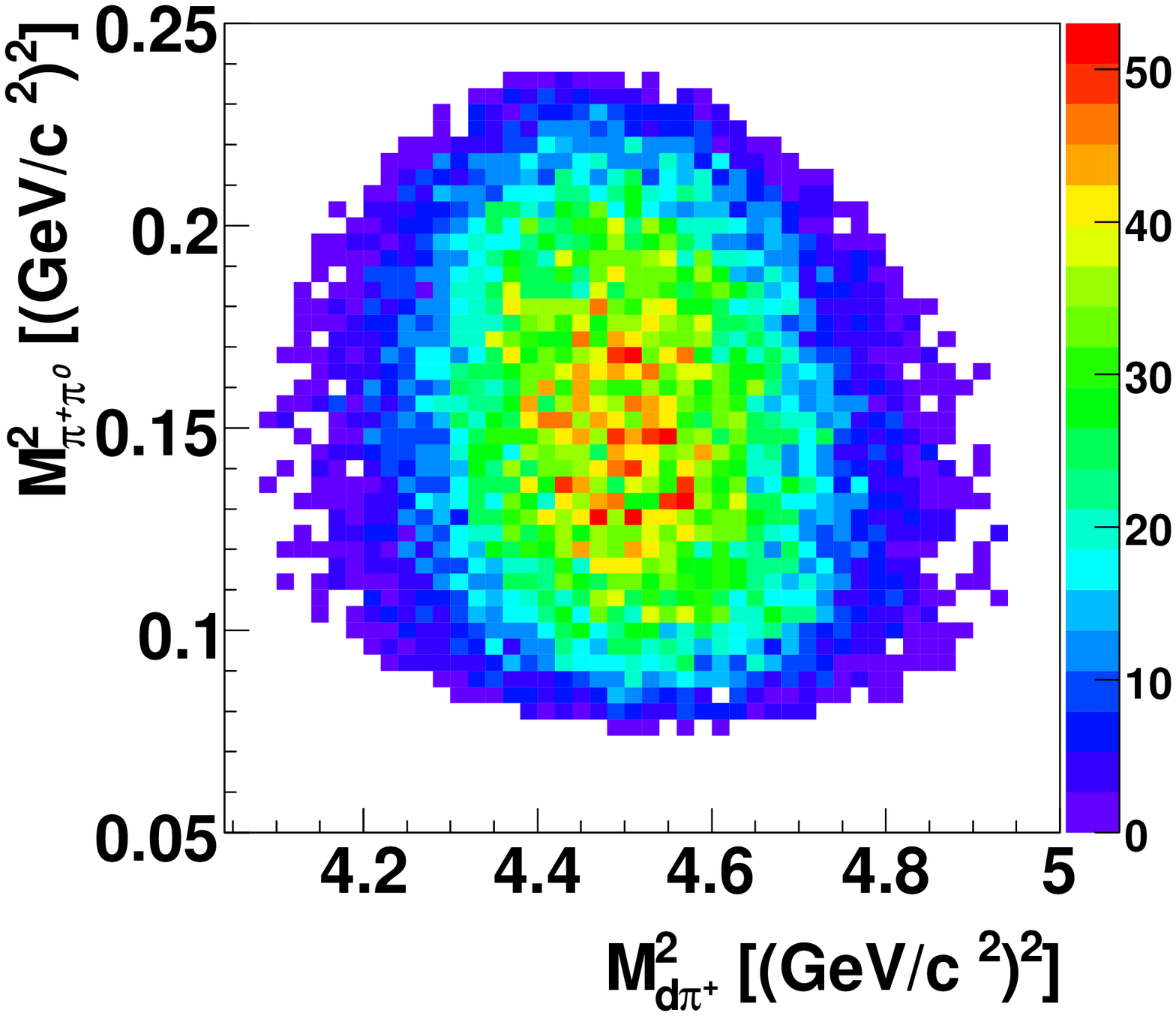}
\includegraphics[width=0.23\textwidth]{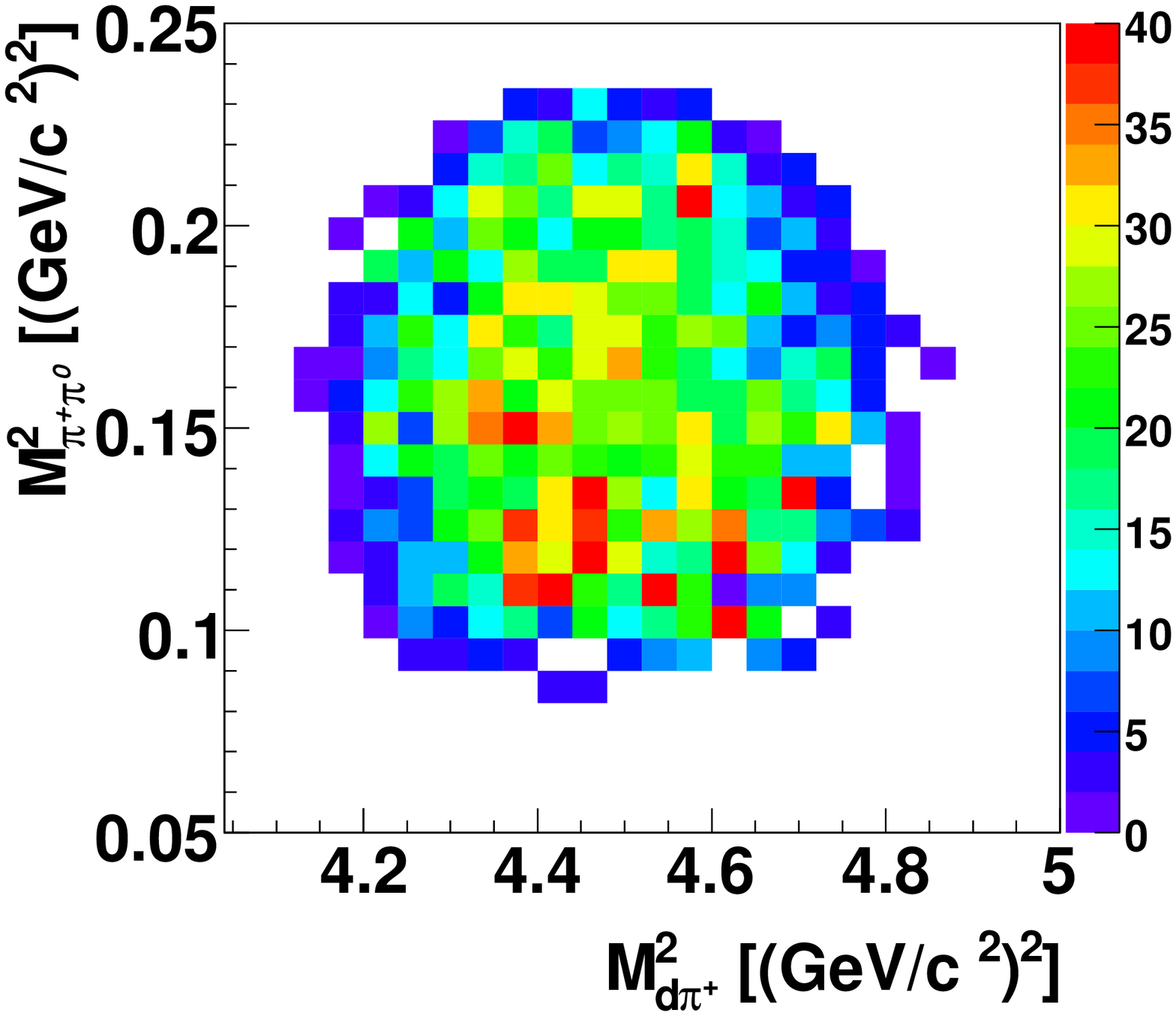}
\includegraphics[width=0.23\textwidth]{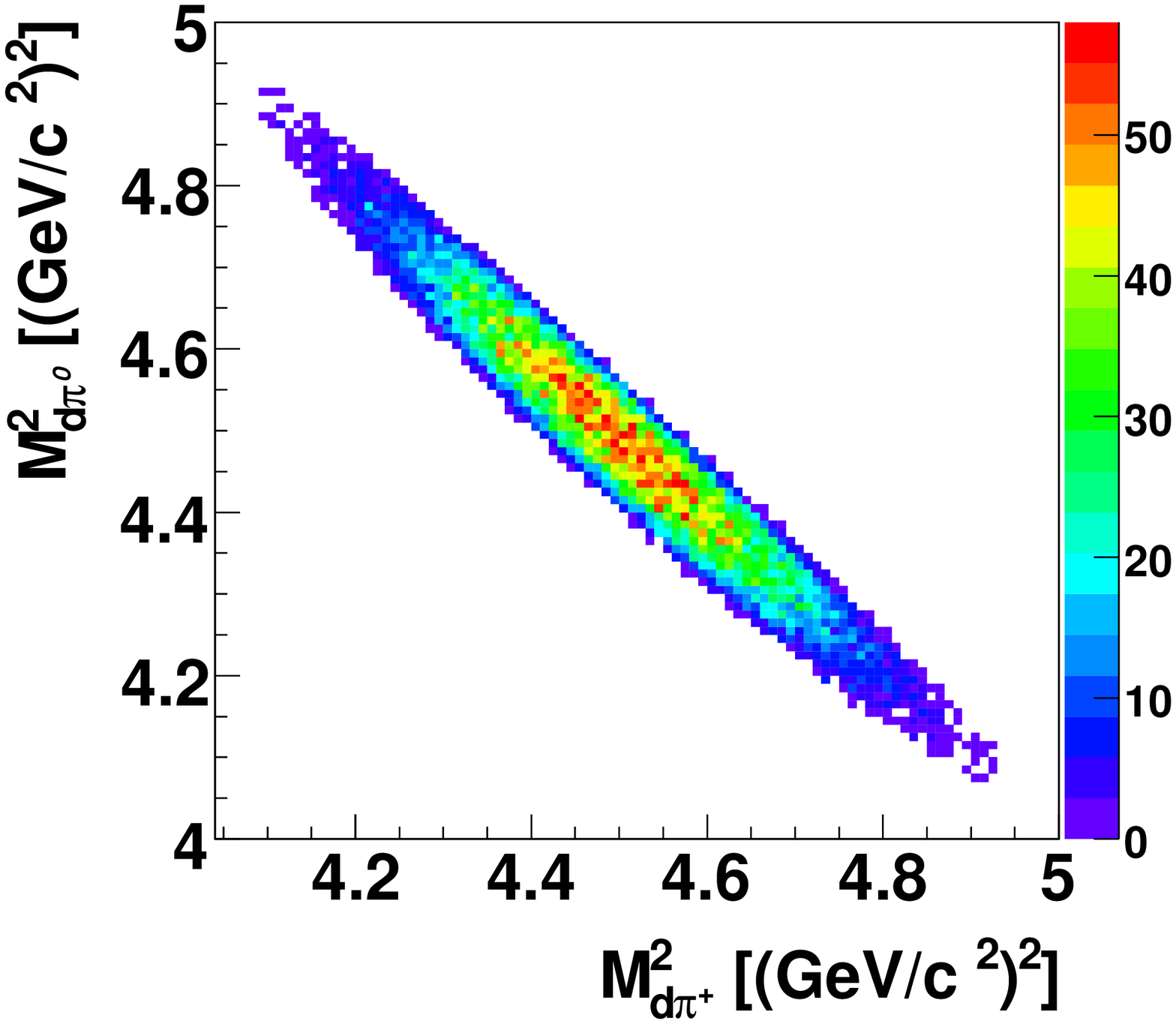}
\includegraphics[width=0.23\textwidth]{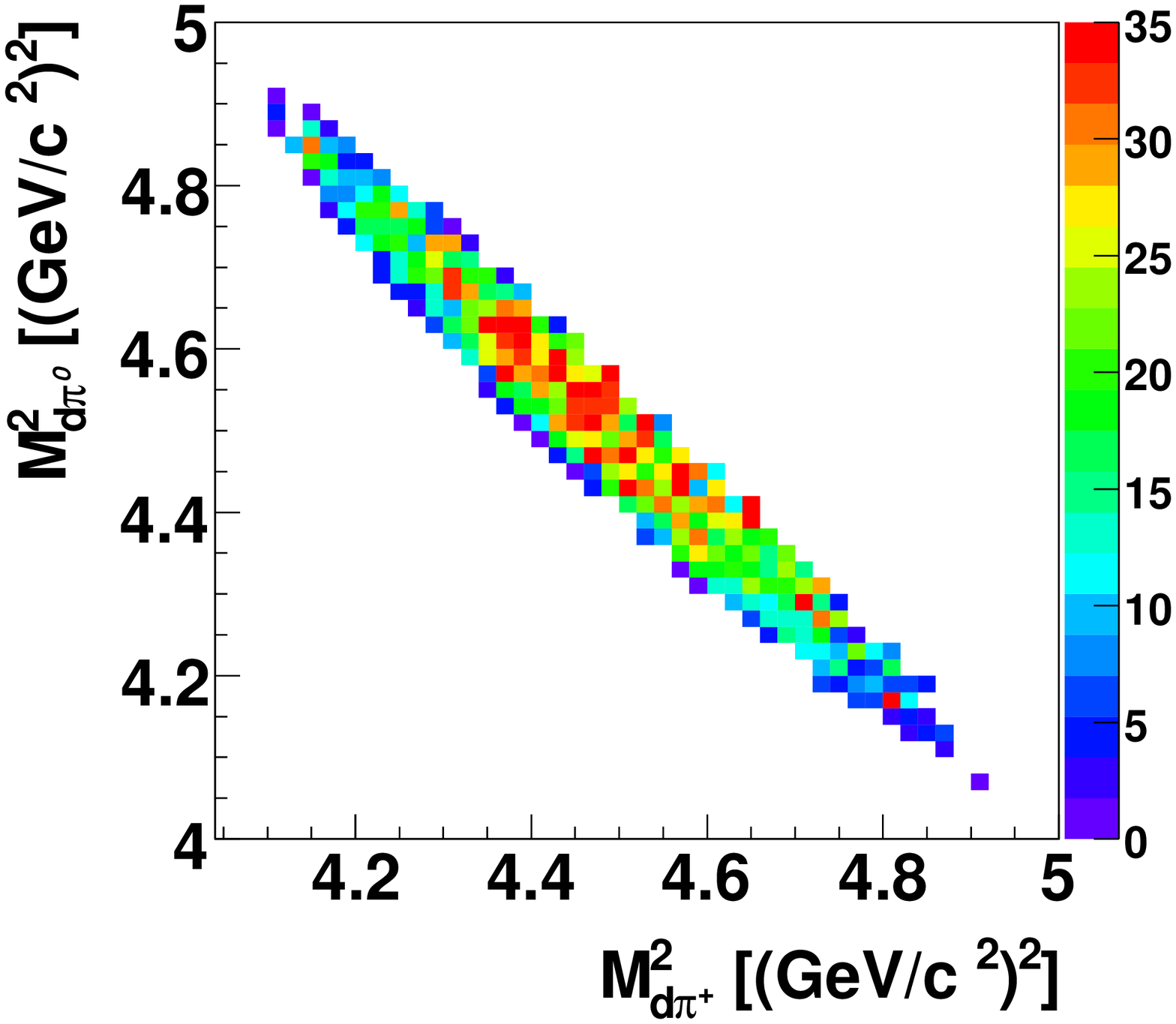}
\caption{ 
  Dalitz plots of the invariant mass squares $M_{d\pi^0}^2$ versus
  $M_{\pi^+\pi^0}^2$ ({\bf top}) and  $M_{d\pi^0}^2$ versus
  $M_{d\pi^+}^2$ ({\bf bottom}) for 
  the $pp \to d\pi^+\pi^0 $ reaction at $T_p$ = 1.1 GeV. {\bf Left}: MC
  simulation of the model description, which corresponds to the solid lines in
  Figs. 4 - 6, {\bf right}: data. 
}
\label{fig1}
\end{center}
\end{figure}

\begin{figure} [t]
\begin{center}
\includegraphics[width=0.23\textwidth]{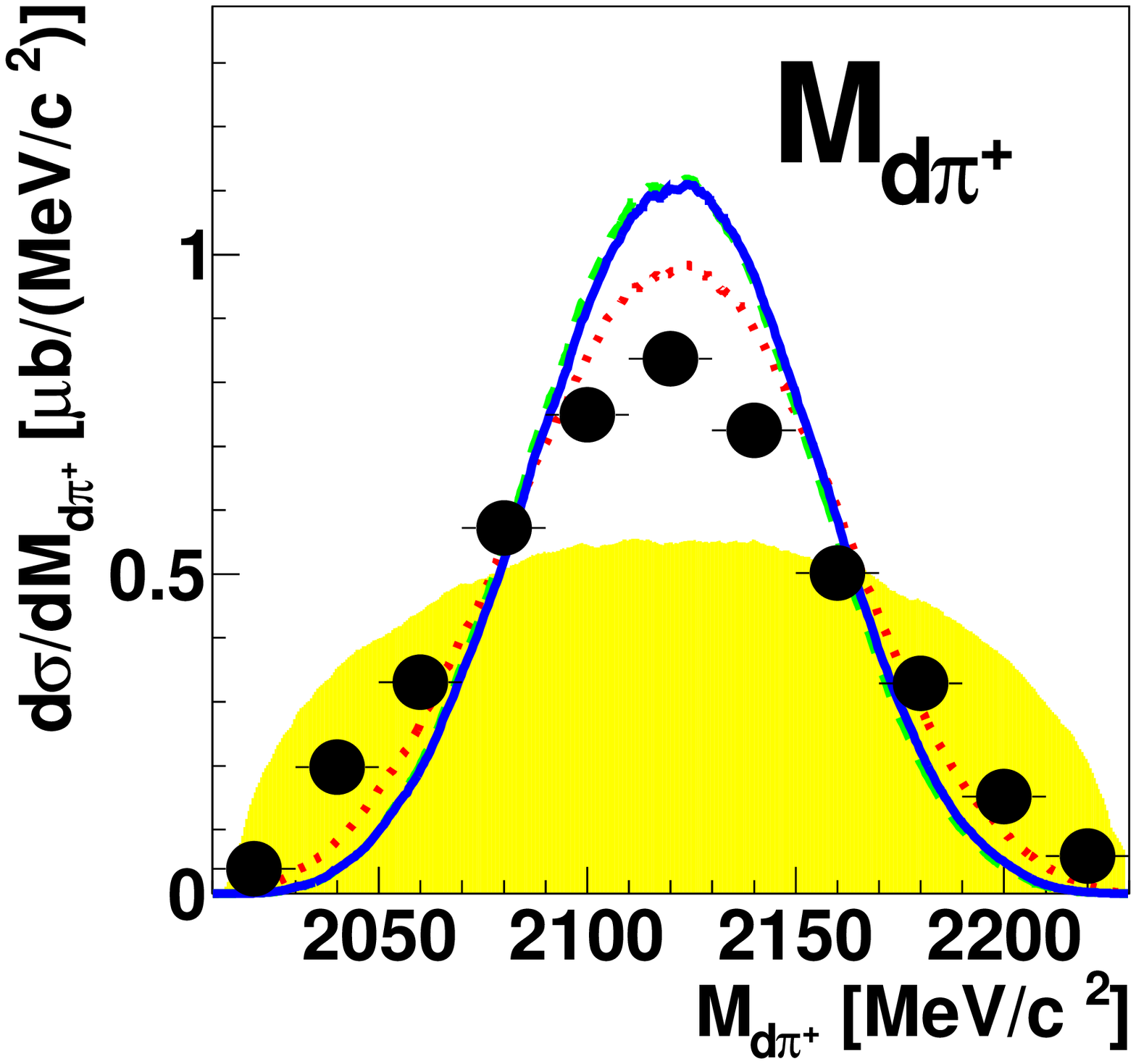}
\includegraphics[width=0.23\textwidth]{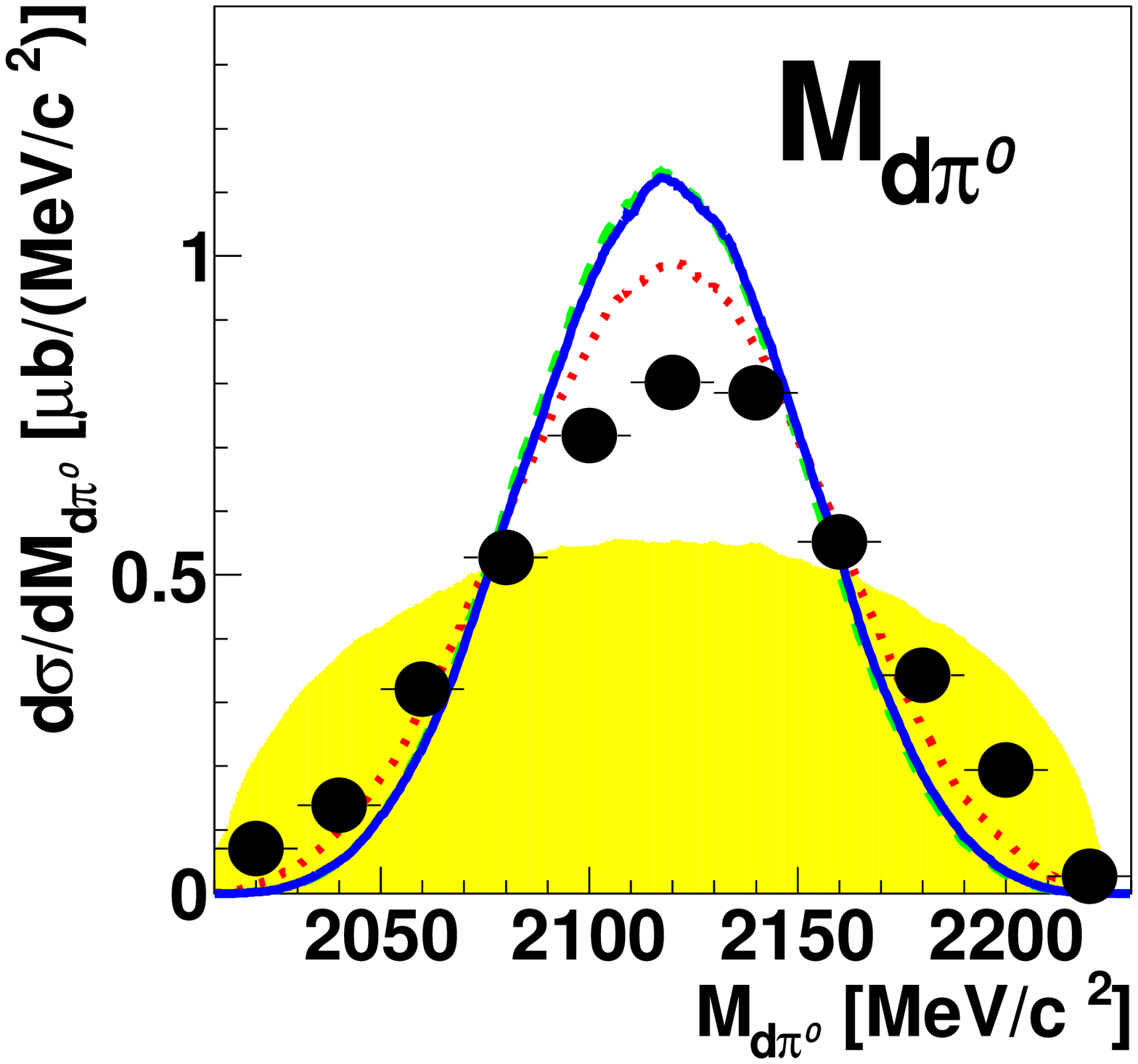}
\includegraphics[width=0.23\textwidth]{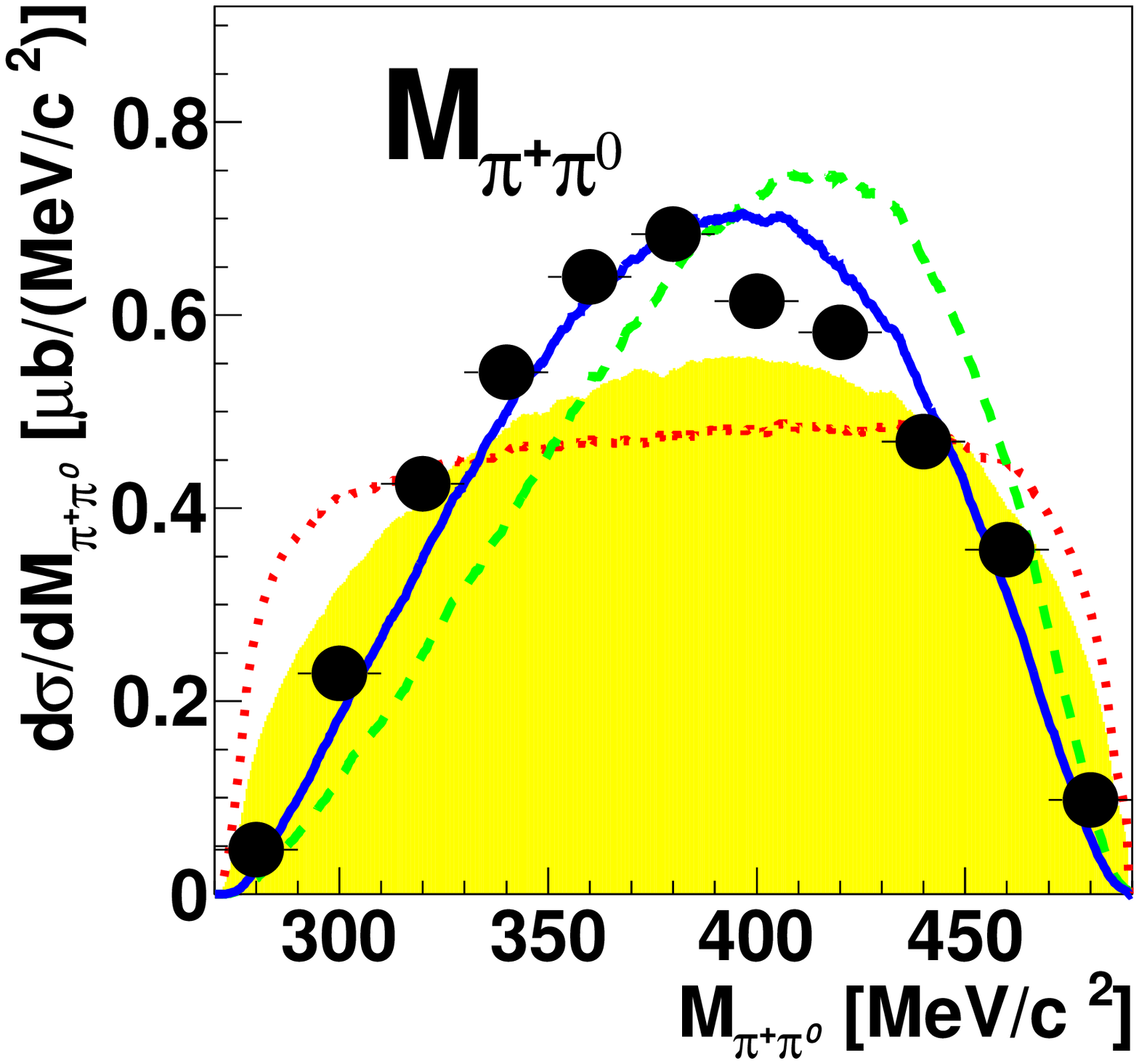}
\includegraphics[width=0.23\textwidth]{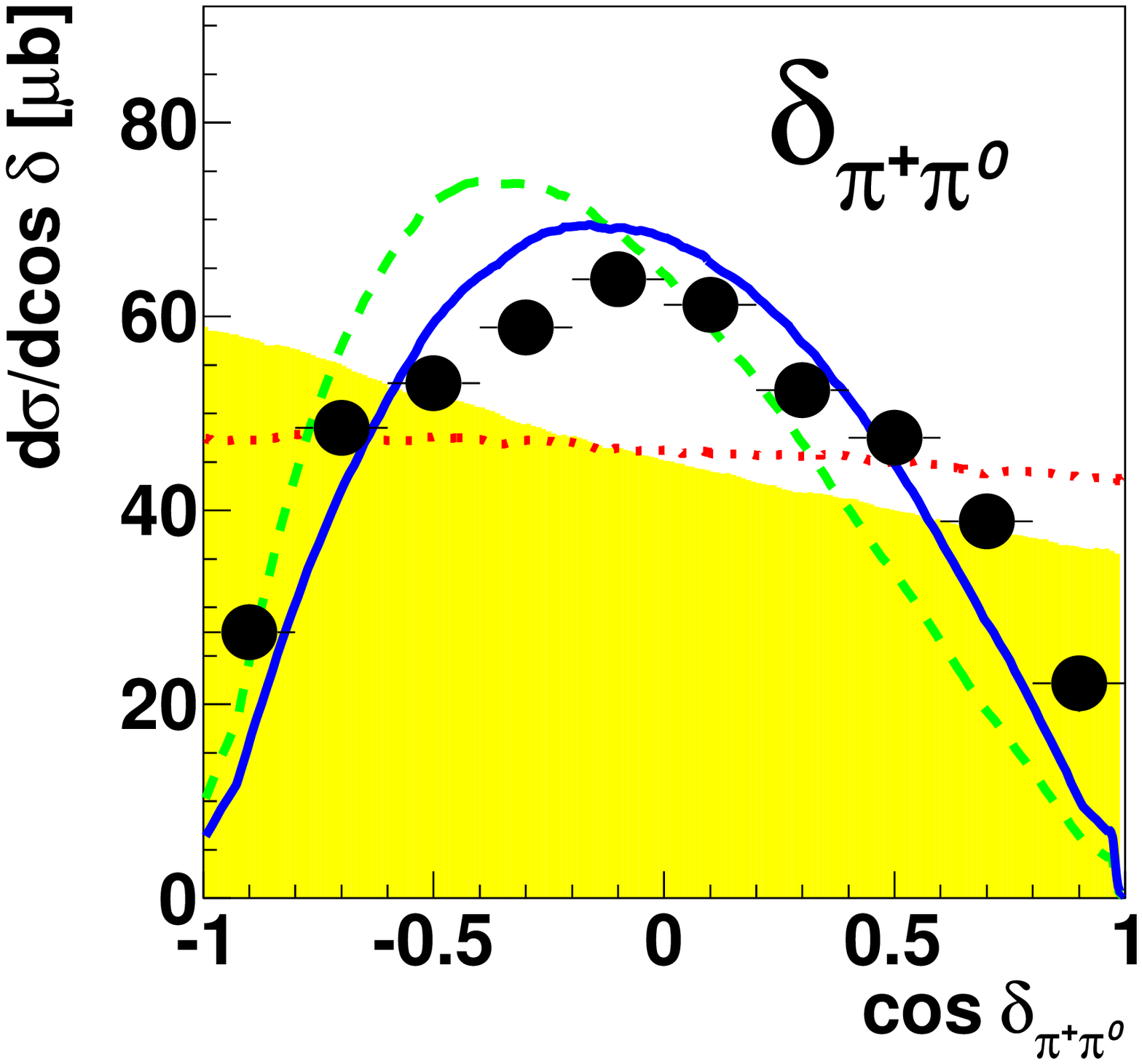}
\caption{ 
  Distributions of the invariant masses $M_{d\pi^+}$ ({\bf top left}), 
  $M_{d\pi^0}$ ({\bf top right}) and $M_{\pi^+\pi^0}$ ({\bf bottom left}) as
  well as of the cm opening angle between the two pions $\delta_{\pi^+\pi^0}$
  ({\bf bottom right}) for 
  the $pp \to d\pi^+\pi^0 $ reaction at $T_p$ = 1.1 GeV. The solid dots
  represent the data of this work. The phase space distributions
  are indicated by the shaded areas. The t-channel $\Delta\Delta$ calculations
  according to Risser and Shuster \cite{ris} are  shown by the dotted curves,
  whereas these calculations supplemented by 
  $\vec{\sigma} * (\vec{k_1}~x~\vec{k_2})$ 
  in the reaction amplitude are given by the solid lines. Dashed curves denote
  calculations for $\rho$ on-shell production.
}
\label{fig1}
\end{center}
\end{figure}

\begin{figure}
\begin{center}
\includegraphics[width=0.23\textwidth]{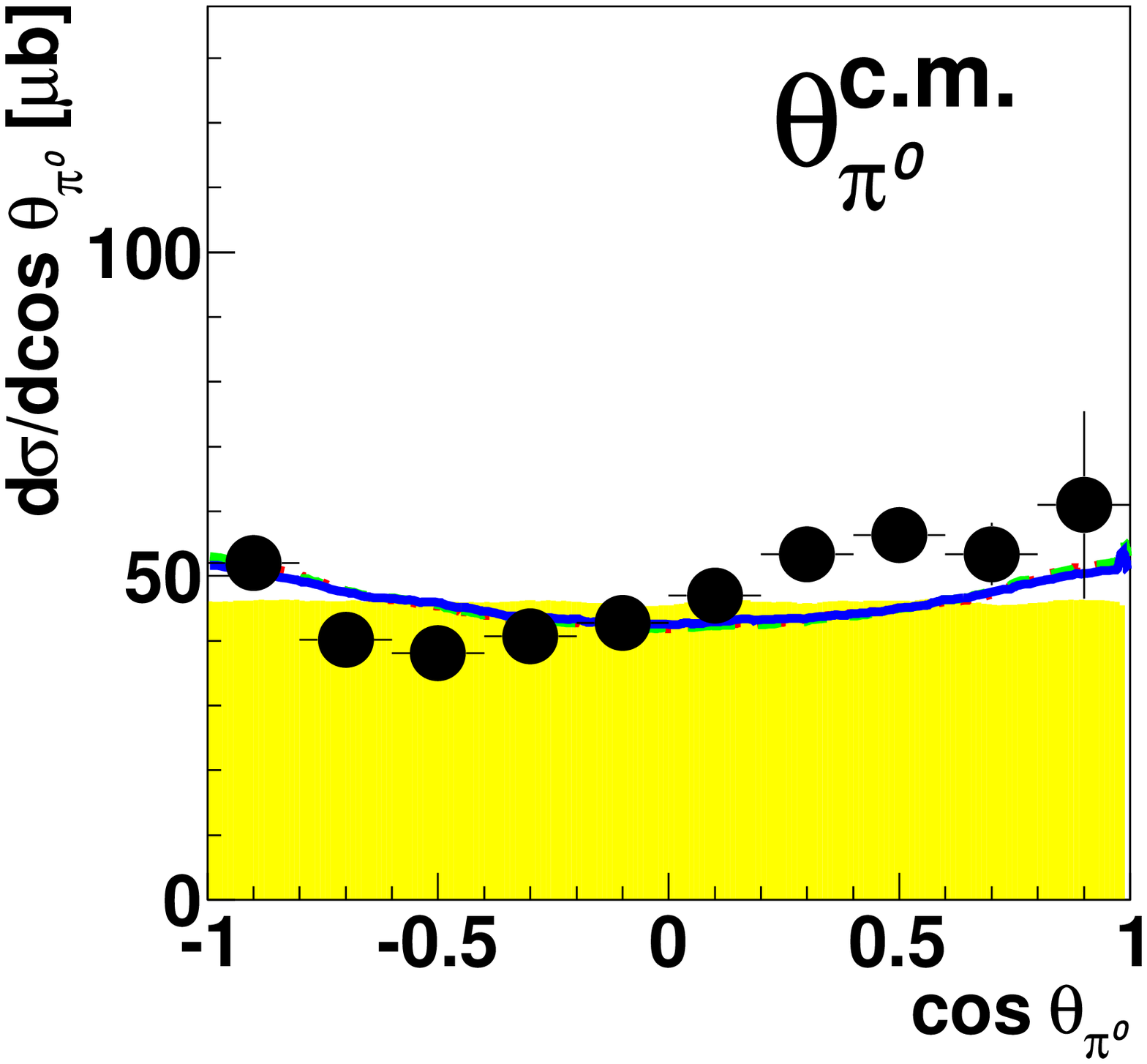}
\includegraphics[width=0.23\textwidth]{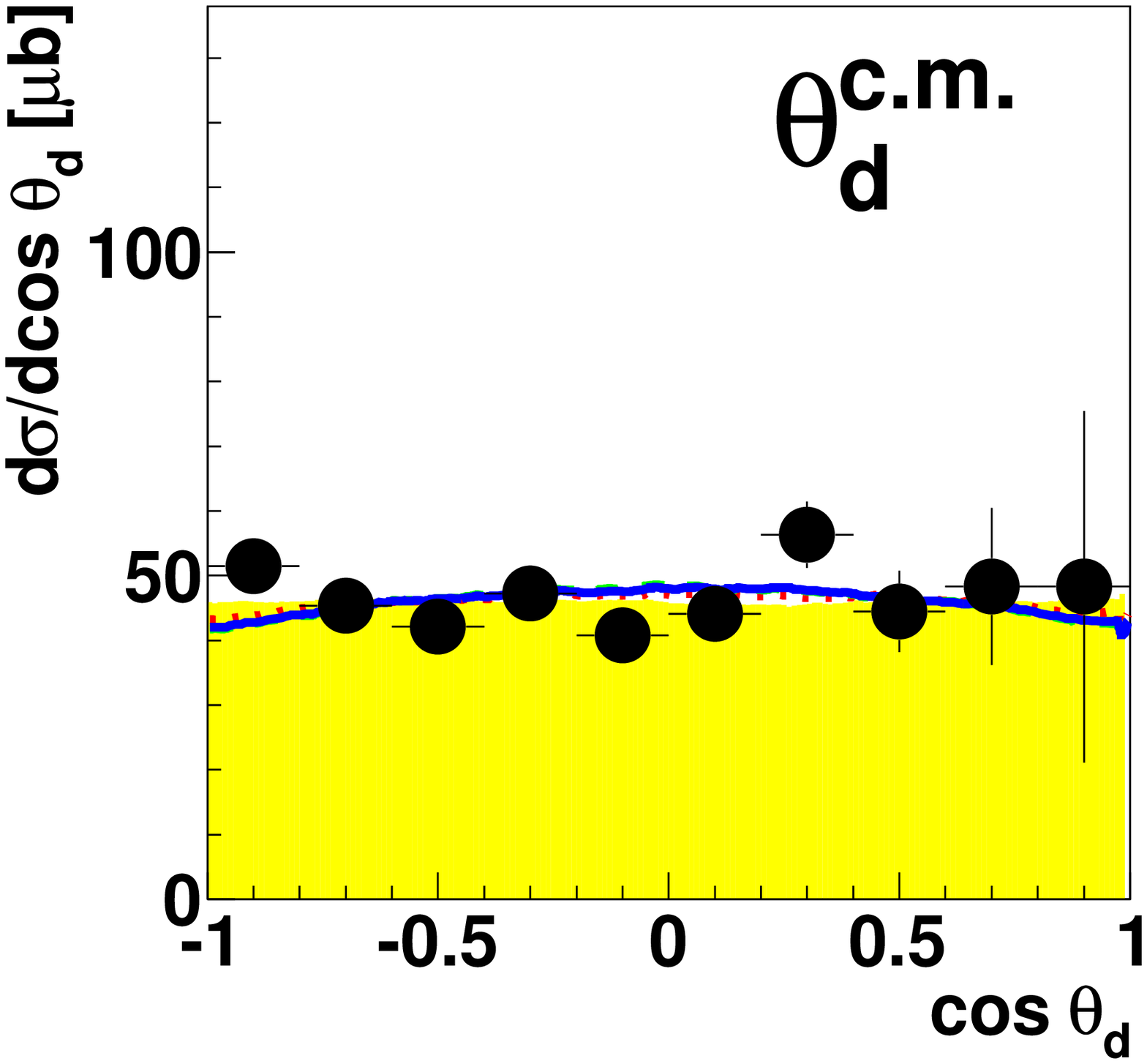}
\caption{ 
  Angular distributions of $\pi^0$ and $d$ in the cm system for 
  the $pp \to d\pi^+\pi^0 $ reaction at $T_p$ = 1.1 GeV. The solid dots
  represent the data from this work. The phase space distribution
  is indicated by the shaded area. For the meaning of the curves see caption
  of Fig. 4. Note that dashed, dotted and solid lines coincide here.
}
\label{fig1}
\end{center}
\end{figure}

\begin{figure} [t]
\begin{center}
\includegraphics[width=0.4\textwidth]{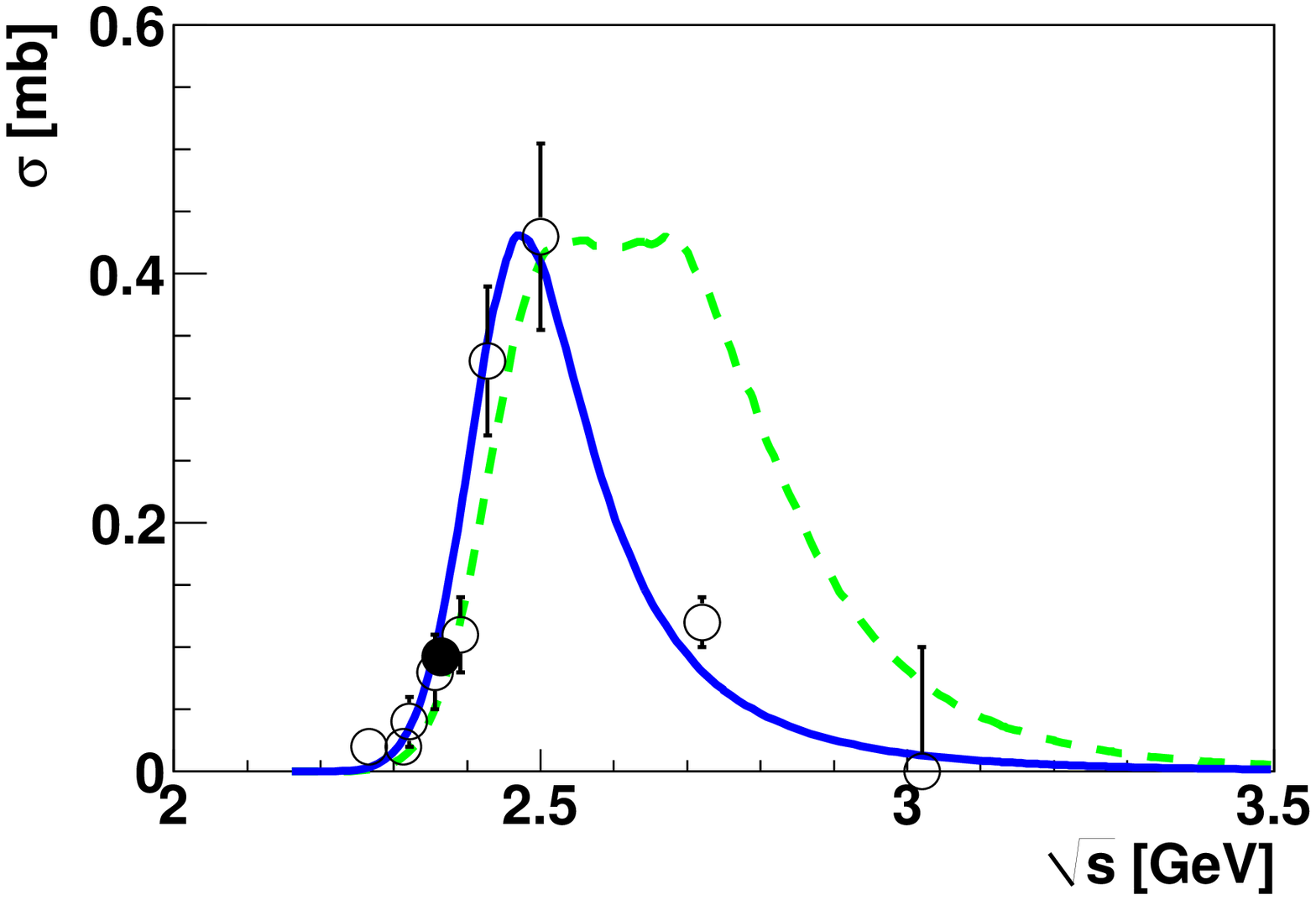}
\caption{
  Energy dependence of the total cross section for 
  the $pp \to d\pi^+\pi^0 $ reaction. The open dots are from previous
  measurements \cite{bys,shim}, the solid dot
  represents the result of this work. For the meaning of the curves see caption
  of Fig. 4. Note that solid and dotted curves coincide here.
}
\label{fig1}
\end{center}
\end{figure}

\section{Results and Discussion}

Results from the measurements are shown in Figs. 3 - 6. Fig. 3 displays the
Dalitz plots of $M_{d\pi^0}^2$ versus $M_{\pi^+\pi^0}^2$ as well as of
$M_{d\pi^0}^2$ versus  $M_{d\pi^+}^2$. In the data there is no evidence for
the presence of the ABC effect, {\it i. e.} for a $\pi\pi$ low-mass
enhancement. Nevertheless the data deviate substantially from phase space, in
particular in the region of the $\Delta$ excitation in $d\pi^0$ and $d\pi^+$
systems. Since we observe this excitation simultaneously in both systems as
revealed in the respective Dalitz plot and especially in its projections, we
see here evidence for the $\Delta\Delta$ excitation in the intermediate system
--- as it is also borne out by comparison to the Monte-Carlo (MC) simulation
of the $t$-channel $\Delta\Delta$ excitation mechanism. 

In its simplest form given by the pioneering ansatz of Risser
and Shuster \cite{ris} the  $t$-channel $\Delta\Delta$ excitation (Fig. 1) is
described by a $t$-channel pion 
exchange followed by two $\Delta$ propagators and the condition that the two
nucleons are constrained in their phase space by the deuteron boundstate
condition. In the isoscalar $\pi\pi$ channel,  where the ABC effect shows up,
the two pion-nucleon p-waves resulting from the decay of the two $\Delta$s
couple either to relative s-waves ($\sigma$-channel) or d-waves between the two
pions. Due to Bose symmetry the isovector $\pi^+\pi^0$ system must be in
relative p-wave to each other ($\rho$ channel), {\it i. e.} the two
pion-nucleon p-waves emerging from the decay of the two $\Delta$ states lead
also to a relative p-wave in the $\pi\pi$ system. This is 
accomplished best if associated with a nucleon spinflip. Hence the
isovector-channel operator $\vec{\sigma} * (\vec{k_1}~x~\vec{k_2})$ as given
in Refs. \cite{alv,luis}, where $\vec{\sigma}$ denotes the Pauli nucleon-spin
operator and $\vec{k_1}$ and $\vec{k_2}$ are the momenta of the outgoing
pions, should be the appropriate operator for describing the isovector
$\pi\pi$ production in $pp \to d\pi^+\pi^0$. 

In Figs. 4 - 6  $t$-channel $\Delta\Delta$ calculations are shown. The dotted
lines denote calculations, where the isovector-channel operator
is omitted ({\it i. e.} as in the  $t$-channel calculations for the ABC
effect in  the isoscalar channel). The solid lines  -- and the MC simulations
for the Dalitz 
plots in Fig. 3 --  denote calculations, where the isovector-channel
operator is applied. The dashed curves finally assume that in the exit channel
the $\pi\pi$ final state interaction leads to the formation of a real $\rho$
meson, {\it i. e.}, these calculations include a Breit-Wigner term (propagator)
for the $\rho$ meson in addition to the vector-isovector operator. Whereas the
latter modification has only a minor impact on the observables, the inclusion
of the isovector-channel operator is striking and essential for the
description of the data --  in particular for the $\pi\pi$ invariant mass and
opening angle distributions. All calculations shown in Figs. 4 - 6 have been
normalized in absolute scale to the observed total cross section.

The observed distributions of the invariant masses $M_{d\pi^+}$ and $M_{d\pi^0}$
exhibit clearly the simultaneous excitation of two $\Delta$ resonances. Simple
t-channel $\Delta\Delta$ calculations without isovector-channel operator
(dotted curves in Figs. 4 - 6) reproduce both $M_{d\pi^+}$ and $M_{d\pi^0}$
spectra quite well, however, not the $M_{\pi^+\pi^0}$ spectrum, where they
exhibit both a low-mass and a high-mass enhancement -- {\it i. e.} the
classically predicted ABC effect for isoscalar channels \cite{ris}. Both these
features are not observed in our data. These calculations also fail to
describe the observed distribution of the opening angle $\delta_{\pi^+\pi^0}$
between the two pions in the center-of- mass (cm) system, see Fig. 4. In
contrast, calculations including the isovector channel operator provide a
reasonable description of all data (solid lines in Figs. 4 - 6). In particular
the $\delta_{\pi^+\pi^0}$ distribution favors the $sin \delta_{\pi^+\pi^0}$
shape predicted by the isovector-channel operator. The further inclusion of a
Breit-Wigner term for the production of a real $\rho$ meson in the final state
(dashed curves) worsens the agreement with the data substantially. 

Fig. 5 displays the angular distributions of $d$ and $\pi^0$ in the
cm system. Both distributions are compatible with isotropy and also in
accordance with the $t$-channel predictions. Since in this experiment the
$\pi^+$ particles could be identified uniquely only in the forward detector
together with the deuterons, the $\pi^0$ particles have been restricted by
kinematics to the backward hemisphere in the cm system. Hence the
$\Theta_{\pi^0}^{cm}$ distribution is not measured over the full angular
range. However, this means no loss of information, since we have identical
particles in the initial channel with the consequence that the cm angular
distributions have to be symmetric about 90$^{\circ}$.

In Fig. 6 finally the energy dependence of the total cross section is
plotted. The open dots give the results from previous bubble chamber
measurements \cite{shim,bys}. The solid dot, which is compatible to the
previous data within uncertainties, gives the result of this work. The drawn
curves represent 
$t$-channel calculations in the definitions given above. They have been
normalized in absolute scale to the data. Solid and dotted curves coincide in
the figure. Their energy dependence is given primarily by the $\Delta$
propagators, whereas the $k^2$ dependence of the reaction amplitude due to the
pion double p-waves from the decay of the two $\Delta$s is counteracted by the
$q^{-2}$ dependence from the pion propagator. This
results in a resonance-like structure with a maximum at twice the $\Delta$
mass and a width of about twice the $\Delta$ width. These calculations give a
good account of the experimentally observed energy dependence. In contrast,
the calculation assuming the production of a real $\rho$ meson in the 
exit channel (dashed curve) is far from the experimentally observed energy
dependence for $\sqrt{s} >$ 2.5 GeV. 

\section{Conclusions}

The first exclusive measurements of the double-pionic fusion reaction to an 
isovector $\pi\pi$ channel provide differential cross sections, which are in
good agreement with a conventional $t$-channel $\Delta\Delta$ excitation
in the intermediate state - though small contributions from other processes
may not be excluded. The isovector $\pi^+\pi^0$ channel exhibits no low-mass
enhancement, {\it i. e.} no ABC effect, as indeed expected from the Bose
symmetry in the $\pi\pi$ system, which prohibits relative s-waves between
$\pi^+$ and $\pi^0$. The fact that also the experimentally observed energy
dependence of the total cross section is well reproduced by the $t$-channel
$\Delta\Delta$ process adds further
confidence to the understanding of the double-pionic fusion process in isovector
channels. Moreover, it provides the possibility to reliably predict the
expected size of the conventional $t$-channel $\Delta\Delta$ process in the
$pn \to d\pi^+\pi^-$ and  $pn \to d\pi^0\pi^0$ reactions by use of isospin
relations as demonstrated in Ref. \cite{MBa}.

\section{Acknowledgments}

We acknowledge valuable discussions with L. Alvarez-Ruso, Ch. Hanhart, E. Oset
and C. Wilkin. This work has been supported by BMBF
(06TU261), Forschungszentrum J\"ulich (COSY-FFE) and  
DFG (Europ. Graduiertenkolleg 683). We also acknowledge the
support from 
the European Community-Research Infrastructure Activity under FP6
"Structuring the European Research Area" programme (Hadron Physics, contract
number RII3-CT-2004-506078).

\section{Important Note}

This is an update of the previous arXiv version published in Physics Letters B
\cite{FK}. The update is based on a reanalysis \cite{fk} of the data for the
$pp\rightarrow d \pi^+ \pi^0$ reaction, where it was discovered that the
originally published value for the cross section was too low by a factor of
about two. This corrected value has been published in an erratum \cite{err}.

The differential distributions obtained in this reanalysis and
exhibited in Figs. 3 - 5  of this updated version did not
change significantly - with the obvious exception in the absolute scale.


\begin{thebibliography}{9}
\bibitem{abc} N. E. Booth, A. Abashian, K. M. Crowe, {\em Phys. Rev. Lett.}
  {\bf 7}, 35 (1961) ; {\bf 5} (1960) 258; {\em Phys. Rev.} {\bf C132}, 2296ff
  (1963)  
\bibitem{ban} J. Banaigs {\it et al.}, {\it Nucl. Phys.} {\bf B67}, 1 (1973)
\bibitem{wur} see; e.g., R. Wurzinger  {\it et al.}, {\it Phys. Lett.} {\bf
    B445}, 423 (1999); for a review see A. Codino and F. Plouin, LNS/Ph/94-06 
\bibitem{col}  F. Plouin, P. Fleury, C. Wilkin, {\em Phys. Rev. Lett.} {\bf
    65}, 690 (1990) 
\bibitem{ris} T. Risser and M. D. Shuster, {\it Phys. Lett.} {\bf 43B}, 68
  (1973)  
\bibitem{anj} J. C. Anjos, D. Levy, A. Santoro, {\it Nucl. Phys.} {\bf B67}, 37
  (1973) 
\bibitem{gar} see, e.g., A. Gardestig, G. F\"aldt, C. Wilkin {\it Phys. Rev.}
  {\bf C59},2608 (1999) and {\it Phys. Lett.} {\bf B421}, 41 (1998);
  C. A. Mosbacher, F. Osterfeld, arXiv: nucl-th/990364
\bibitem{alv} L. Alvarez-Ruso, {\em Phys. Lett.} {\bf B452}, 207(1999); PhD
  thesis, Univ. Valencia 1999  
\bibitem{bash} M. Bashkanov {\it et al.}, {\em Phys. Lett.} {\bf B637} (2006)
  223; arXiv: nucl-ex/0508011
\bibitem{MBa} M. Bashkanov {\it et al.},  {\em Phys. Rev. Lett.} {\bf 102}
  (2009)  052301; arXiv: 0806.4942 [nucl-ex] 
\bibitem{hcl} H. Clement {\it et al.}, {\em Prog. Part. Nucl. Phys..} {\bf 61}
  276 (2008); arXiv: 0712.4125 [nucl-ex] 
\bibitem{sk} S. Keleta {\it et al.}, {\em Nucl. Phys.} {\bf A825} (2009) 71
\bibitem{panic} M. Bashkanov {\it et al.},  {\em Proc. PANIC08 (Elsevier,
    eds. I. Tserruya, A. Gal, D. Ashery)} (2009)  239; arXiv:
  0906.2328 [nucl-ex]
\bibitem{iso} T. Skorodko {\it et al.}, {\em Phys. Lett.} {\bf B 6679} (2009)
  30; arXiv: 0906.3087 [nucl-ex]
\bibitem{luis} L. Alvarez-Ruso, E. Oset, E. Hernandez, {\em Nucl. Phys.} {\bf
    A633} (1998) 519 and priv. comm.
\bibitem{shim} F. Shimizu {\it et al.}, {\em Nucl. Phys.} {\bf
    A386} (1982) 571
\bibitem{bys} J. Bystricky {\it et al.}, {\em J. Physique} {\bf
    48} (1987) 1901 and references therein
\bibitem{barg} Chr. Bargholtz {\it et al.}, {\em Nucl. Instr. Meth.} {\bf
    A594} (2008) 339
\bibitem{MB} M. Bashkanov, PhD thesis, Univ. T\"ubingen 2006
\bibitem{FK} F. Kren {\it et al.}, {\em Phys. Lett.} {\bf B684} (2010) 110
\bibitem{fk} F. Kren, Dissertation, Univ. T\"ubingen 2010:
http://tobias-lib.uni-tuebingen.de/volltexte/2010/5034/pdf/doki13.pdf
\bibitem{err} F. Kren {\it et al.}, {\em Phys. Lett.} {\bf B702} (2011) 312
\end{thebibliography}
\end{document}